\documentclass[aps,pra,10pt,twocolumn,showpacs,superscriptaddress]{revtex4-1}

\usepackage{amsmath}
\usepackage{bm}
\usepackage{latexsym}
\usepackage{amssymb} 
\usepackage{amsthm} 
\usepackage{times} 

\begin{document}


\title{Interference Visibility as a Witness of Entanglement and Quantum Correlation}


\author{Lin Zhang}

\affiliation{ Institute of Mathematics, \\
Hangzhou Dianzi University, Hangzhou 310018, PR~China}

\author{Arun Kumar Pati}
\affiliation{Quantum Information and Computation Group,\\
Harish-Chandra Research Institute, Chhatnag Road, Jhunsi, Allahabad
211 019, India}
\affiliation{Department of Mathematics\\
Zhejiang University, Hangzhou 310027, PR~ China}

\author{Junde Wu}
\affiliation{Department of Mathematics\\
Zhejiang University, Hangzhou 310027, PR~ China}

\date{\today} 

\begin{abstract}
In quantum information and communication one looks for the non-classical features like
interference and quantum correlations to harness the true
power of composite systems. We show how the concept akin to
interference is, in fact, intertwined in a quantitative manner to entanglement and
quantum correlation. In particular, we prove that the difference in
the squared visibility for a density operator before
and after a complete measurement, averaged over all unitary
evolutions, is directly related to the quantum correlation measure based on the
measurement disturbance. For pure and mixed bipartite states the unitary average of the
squared visibility is related to entanglement measure. This may
constitute direct detection of entanglement and quantum correlations with  quantum
interference setups. Furthermore, we prove that for a fixed
purity of the subsystem state, there is a complementarity relation
between the linear entanglement of formation and the measurement disturbance.
This brings out a quantitative difference between two kinds
of quantum correlations.
\end{abstract}

\maketitle


\section{ Introduction}

Quantification and detection of quantum
entanglement and general quantum correlations are of paramount
importance in the field of quantum information. It was already
realized in the early days of quantum theory that linear
superposition and entanglement are two essential features that
distinguishes the quantum world from the classical world \cite{er,epr}.
The principle of linear superposition is at the heart of most of the
counter intuitive phenomena that we see in quantum mechanics and
this also gives rise to quantum entanglement for multiparticle pure
states. However, for mixed states in addition to entanglement there can be other type
of non-classical correlations beyond entanglement. Among mixed state
entanglement measures there are many, for e.g. the concurrance \cite{woo},
entanglement of formation \cite{chb1,chb2},
relative entropy of entanglement \cite{ved}, logarithmic negativity \cite{vidal} and many more \cite{ho,guh}.
The non-classical correlations that are not based on the separability paradigm can be quantum discord \cite{oz,hv}, work
deficit \cite{jo,sen}, quantumness of correlation \cite{akr},
measurement induced disturbance \cite{luo}, geometric discord \cite{dak},
super discord \cite{sp} (see for example a recent review  \cite{modi}).
The notion of quantum entanglement briefly refers to nonlocal property of a global system
that cannot be emulated by local descriptions (with local operation and classical communication (LOCC)).
On the other hand, the quantum correlation in a broader sense
may be thought of as how much quantumness one cannot access about
the state of the global system by accessing only one subsystem.  The
quantum correlation which is not captured by entanglement measures,
has been a subject of growing interest in recent years. It is
important to propose feasible schemes that can distinguish classical
correlations, entanglement and genuine quantum correlations present in composite
quantum systems. Several proposals have come up in recent years to
detect non-classical correlations. To name few proposals, it has
been suggested that the quantum correlation can be quantified by
measuring the expectation value of a small set of observables on
four copies of the state \cite{giro1}. Even intrinsic quantum
uncertainty for a single observable gives a measures of nonclassical
correlations similar to that of the discord \cite{giro2}. Also,
quantum states with non-zero correlation can lead to a nonzero precision
in the parameter estimation \cite{giro3}.

In interference setup, typically, one considers a particle in a pure
state. If one coherently splits an incident particle and applies
a unitary transformation on one arm of the interferometer, upon
recombing the two paths one sees an interference. The visibility in
the interference depends on $|\langle \Psi|U|\Psi\rangle|^2$.
However, if we have mixed states, then it was not easy to define the
relative phase and the visibility. In an important paper \cite{es},
it was proved that if we send a mixed state through a Mach-Zender
interferometer, such that $\rho \rightarrow \rho' = U\rho
U^{\dagger}$ then the \emph{relative phase} between $\rho$ and
$\rho'$ is given by $\mathrm{Arg} \mathrm{Tr}(\rho U)$ and the
\emph{visibility} is defined as $V = V(\rho, U)= |\mathrm{Tr}(\rho U)|$.
This generalizes the notion of relative phase shift known as the
Pancharatnam phase from pure states \cite{panch} to mixed states.
The relative phase and the visibility for mixed states have been measured in NMR and photon interference experiments
\cite{du,me,anil}. The notion of visibility for the mixed state has
found several applications in recent years. It has been suggested
that using the interferometric setup one can measure linear and
non-linear functions of the density operator \cite{ae} as well as
detect entanglement of unknown density operator which needs
estimation of $(d^2-1)$ parameters and joint measurement on $d$
copies of (transformed) density operator \cite{ekert}. This also
paves the way to define the notion of interference of quantum
channels \cite{dk}. Recently, it has been shown that using the
notion of visibility one can define a new metric for the density
operator along the unitary orbit and this gives a tighter bound on
the quantum speed limit compared to other known bounds \cite{mp}.

Non-classical correlation in the composite system is not only a
resource, but it also plays a major role in understanding the
dynamics of reduced systems of a composite system. Presence of
initial correlations prohibit the description of the reduced dynamics to
be completely positive map \cite{pp,ra,gl,ar,ps,jss}.  In this
regard methods for the detection of quantum correlation in
system-environment has been proposed and it has been shown that the
amount to which the evolution of the reduced state differs from the
evolution of the local dephased reduced state can detect the initial
correlation \cite{gb}.

In this paper, we explore if it is possible to quantify the
non-classical correlation and entanglement using the concepts of
quantum interferometry for mixed states. We recourse to the notion
of interference of mixed states and prove that the difference in the
unitary average of the mixed state (squared) visibility of the
density operator and its dephased counter part is directly related
to the quantum correlation measure which is defined via the
measurement disturbance.  Then, we show that our method of detecting 
quantum correlation is also applicable for
noisy measurements. For pure bipartite states the average of the difference in the squared 
visibility before and after the measurement is
related to the concurrence.  Furthermore, for pure bipartite states,
we prove that under local unitary evolution of one subsystem, the
unitary average of the squared visibility gives the entanglement.
For mixed states, using the convex roof construction, we give a new
lower bound for the linear entropy version of the entanglement of formation in terms of the
average visibility.
In addition, we prove that for a fixed
purity of the subsystem state, there is a complementarity relation
between the linear entanglement of formation and the measurement disturbance.
Possibly, this suggests a new difference between two kinds
of quantum correlations, hitherto unnoticed. Our result makes a direct connect between  two
fundamental ingredients of quantum theory, namely, the quantum
interference and the quantum correlation. This may constitute direct
detection of entanglement including other non-classical correlations
with quantum interference setup. Our method can be tested in
interference experiments and the presence of non-classical
correlations can be revealed.

The paper is organized as follows. In section II, we show that the difference
in the average of the interference visibility for the density operator before
and after dephasing is directly related to the measurement disturbance.
In section III, we show how our method can detect quantum correlation with the noisy
measurements.
In section IV, we show that for pure bipartite states, having access to one subsystem, the
unitary average of the visibility gives the concurrence. For mixed
states we show that the entanglement of formation is bounded
from above by the average of the interference visibility.
In section V, we show that for a fixed purity of the subsystem state,
there is a complementarity between entanglement of formation and the
measurement disturbance.
Finally, we conclude in section VI.



\section{ Quantum correlation with visibility}

Given a bipartite state
$\rho_{AB}$ on ${\cal H}_A \otimes {\cal H}_B$, in general, it
contains classical correlation \cite{hv}, non-classical correlation
such as the entanglement \cite{ho} and the statistical correlation of
quantum origin that is not captured by the entanglement
\cite{oz,hv,jo,sen,akr,luo,sp}. We will use the quantum correlation
measure based on the measurement disturbance \cite{luo}. In quantum
mechanics, a projective measurement on a quantum system usually
disturbs the state and looses its quantumness. Similarly, for a
composite system, a measurement carried out on one part (or both)
usually disturbs the state thereby destroying the quantum
correlation present in the system. If $\{\Pi_i^A \}$ and $\{\Pi_i^B
\}$ are complete one-dimensional orthogonal projectors in ${\cal
H}_A$ and ${\cal H}_B$, then after the measurement process the
density operator transforms as
\begin{align}
\rho_{AB} \rightarrow
\Phi(\rho_{AB})= \sum_{i,j} (\Pi_i^A \otimes \Pi_j^B) \rho_{AB}
(\Pi_i^A \otimes \Pi_j^B).
\end{align}
The measurement disturbance based
quantum correlation is defined as \cite{luo}
\begin{eqnarray}
Q(\rho_{AB}) := \|\rho_{AB} - \Phi(\rho_{AB})\|_\mathrm{HS},
\end{eqnarray}
where $\|A\|_\mathrm{HS} = \sqrt{\mathrm{Tr}(A^{\dagger} A)}$ is the
Hilbert-Schmidt norm for an operator $A$.
Physically, this tells us the extent to which the measurement
operator disturbs the composite state and gives a measure of
quantumness based on the distance between the classical state
closest to the original bipartite state. For classical state the
quantum correlation measure is zero. One can also use a quantum
correlation based on measurement disturbance where projective
measurement is performed on one subsystem.

Consider a bipartite
state $\rho_{AB}$ that undergoes a unitary evolution $\rho_{AB}
\rightarrow U\rho_{AB} U^{\dagger}$. If one interferes the original
and the unitary transformed density operator the visibility is given
by
\begin{eqnarray}
V^2(\rho_{AB}, U) = |\mathrm{Tr}(\rho_{AB} U)|^2.
\end{eqnarray}
Now, we perform local projective measurements on $\rho_{AB}$ and
evolve the state $\Phi(\rho_{AB})$ under the unitary evolution.
Since the measurement process destroys the quantumness, it is natural that if we interfere
$\Phi(\rho_{AB})$ with $U\Phi(\rho_{AB}) U^{\dagger}$, then there
will be some change in the interference pattern, both in terms of
the relative phase shift and the visibility. Now, the important question we ask is 
for generic
unitary operators how much change does occur in the visibility of the
interference. If we consider the unitary operator as a unitary
matrix which has been drawn from an appropriate random matrix
ensemble over the unitary group $\mathrm{U}(d)$, then the unitary
average of change in the visibility is directly related to the
quantum correlation. Throughout our paper, we will use normalized
bi-invariant Haar measure $\mu$ over unitary matrix group
$\mathrm{U}(d)$.\\

\noindent\textbf{Theorem 1.} For any bipartite density operator
$\rho_{AB}$ on ${\cal H}_A \otimes {\cal H}_B$, the difference
between the unitary average of the visibility before and after
measurement is a measure of quantum correlation, i.e.,
\begin{eqnarray}
&&\int_{\mathrm{U}(d_A d_B)}\left[|\mathrm{Tr}(\rho_{AB} U)|^2 -
|\mathrm{Tr}(\Phi(\rho_{AB})
U)|^2\right]d\mu(U)\nonumber\\
&&=\frac1{d_Ad_B}Q(\rho_{AB})^2.
\end{eqnarray}

\begin{proof}
The average of the (squared) visibility of quantum interference for
the mixed states for generic unitary operators can be expressed as
\begin{eqnarray*}
\int_{\mathrm{U}(d)}|\mathrm{Tr}(\rho_{AB} U)|^2d\mu(U) &=&
\mathrm{Tr}\left(\rho^{\otimes 2}_{AB}\int_{\mathrm{U}(d)} U\otimes U^\dagger
d\mu(U)\right) \nonumber\\
&=& \mathrm{Tr}\left(\rho^{\otimes 2}_{AB} M \right),
\end{eqnarray*}
where $M = \int_{\mathrm{U}(d)} U\otimes U^\dagger d\mu(U)$. First we prove that
\begin{eqnarray}\label{eq:uu*}
M =\int_{\mathrm{U}(d)} U\otimes U^\dagger d\mu(U) = \frac{F}{d},
\end{eqnarray}
where $F$ is the swap operator defined as
$F=\sum_{i,j}|ij\rangle\langle ji|$ and $d= d_A d_B$.

First, we note that 
$M$ is self-adjoint by the property of the Haar-measure.
From the fact that $\mathrm{Tr}((A\otimes B) F) =\mathrm{Tr}(AB)$,
we have $\mathrm{Tr}(MF)=d$. Since the Haar-measure is left-regular,
it follows that $(V\otimes I)M(I\otimes V^\dagger) = M$ for all
$V\in\mathrm{U}(d)$. By taking traces over both sides, we have
$\mathrm{Tr}(M) = \mathrm{Tr}(M(V\otimes V^\dagger))$. Now, by
taking integrals over both sides, we have
$\mathrm{Tr}(M)=\mathrm{Tr}(M^2)$. By the Cauchy-Schwartz
inequality, we get
$$
d^2 = (\mathrm{Tr}(MF))^2\leqslant \mathrm{Tr}(M^2)\mathrm{Tr}(F^2)
= d^2\mathrm{Tr}(M),
$$
which implies that $\mathrm{Tr}(M)\geqslant1$. In what follows, we show
that $\mathrm{Tr}(M)=1$. By the definition of $M$, we have
\begin{eqnarray}
\mathrm{Tr}(M) = \langle\langle
I_d|\int_{\mathrm{U}(d)}|U\rangle\rangle\langle\langle U|
d\mu(U)|I_d\rangle\rangle,
\end{eqnarray}
where $|X\rangle\rangle := \sum_{i,j}X_{ij}|ij\rangle$ for a matrix
$X=\sum_{i,j}X_{ij}|i\rangle\langle j|$. Now, define a unital quantum
channel $\Gamma$ as follows:
$$
\Gamma := \int_{\mathrm{U}(d)}\mathrm{Ad}_{U}d\mu(U).
$$
Thus, we have $\Gamma(X)=\mathrm{Tr}(X)\frac{I_d}{d}$.
By the
Choi-Jamio{\l}kowksi isomorphism, it follows that
\begin{eqnarray}
J(\Gamma) &=&
(\Gamma\otimes\mathrm{id})(|I_d\rangle\rangle\langle\langle I_d|)
\\&=&
\int_{\mathrm{U}(d)}|U\rangle\rangle\langle\langle U| d\mu(U).
\end{eqnarray}
For the completely depolarizing channel $\Gamma(X) =
\mathrm{Tr}(X)\frac{I_d}{d}$, we already know that $J(\Gamma) =
\frac1d I_d\otimes I_d$. Then, it follows that
\begin{eqnarray}
\int|U\rangle\rangle\langle\langle U| d\mu(U) = \frac1d I_d\otimes
I_d.
\end{eqnarray}
Finally, we have $\mathrm{Tr}(M)=1$. This indicates that the
Cauchy-Schwartz inequality is saturated, and moreover the saturation
happens if and only if $M=\lambda F$ for $\lambda$ constant. By
taking traces over both sides, we have $\lambda=\frac1d$. The
desired identity is obtained.

Therefore, by \eqref{eq:uu*}, the expression for the visibility of
quantum interference for the mixed states averaged over all unitary
group using the Haar measure is given by
\begin{eqnarray}\label{eq:main}
\int_{\mathrm{U}(d_A d_B)}|\mathrm{Tr}(\rho_{AB} U)|^2d\mu(U) =
\frac1{d_Ad_B}\mathrm{Tr}(\rho^2_{AB}).
\end{eqnarray}
This shows that for generic unitary evolutions of quantum system,
it is the purity of the
density operator that ultimately decides the visibility in the
interference. Since the measurement disturbance $Q(\rho_{AB})^2
= \mathrm{Tr}(\rho^2_{AB}) - \mathrm{Tr}(\Phi(\rho_{AB})^2)$, we
have the proof.
\end{proof}

Therefore, the unitary average of the difference in
the (squared) visibility before and after the measurement performed
on the composite system is given by the quantum correlation based on
measurement disturbance.

In our proof of Theorem 1, we have used global
unitary and have shown that the unitary average of the interference
visibility before and after a complete measurement is related to
quantum correlation. However, we can also have the same result when
the interference visibility is obtained under local unitary and
average over local unitary groups.
Detection of quantum entanglement and quantum correlation using local operations is
important when we do not have access to the whole system to reveal
these correlations present in the composite system.
Therefore, our method can help in detection of
quantum correlation with access to local subsystems only. In fact, we can prove that 
for any bipartite state $\rho_{AB}$, the unitary average of visibility under
local unitary satisfies
\begin{eqnarray}
&&\int_{\mathrm{U}(d_A)}\int_{\mathrm{U}(d_B)}|\mathrm{Tr}(\rho_{AB} (U \otimes V))|^2d\mu(U)d\mu(V) \nonumber\\
&&= \frac1{d_Ad_B}\mathrm{Tr}(\rho^2_{AB}).
\end{eqnarray}
To prove this consider the spectral decomposition of $\rho_{AB}$,
with $\rho_{AB} = \sum_j \lambda_j |\Phi_j\rangle\langle \Phi_j|$,
where $\lambda_j$'s and $|\Phi_j\rangle$'s are the eigenvalues and
the eigenvectors of $\rho_{AB}$, respectively. Now we see that there
exist $d_B\times d_A$ matrices $Y_j$ such that $|\Phi_j\rangle =
|Y_j\rangle\rangle$. This suggests that we can write
\begin{eqnarray*}
&&|\mathrm{Tr}(\rho_{AB}(U\otimes V))|^2 =\sum_{i,j}\lambda_i\lambda_j\\
&& \times\mathrm{Tr}\left[(Y^\dagger_i\otimes Y^\dagger_j) (U\otimes
U^\dagger) (Y_i\otimes Y_j)(V^\mathtt{T}\otimes
(V^\mathtt{T})^\dagger)\right].
\end{eqnarray*}
Thus, we have
\begin{eqnarray*}
&&\int_{\mathrm{U}(d_A)}\int_{\mathrm{U}(d_B)} |\mathrm{Tr}(\rho_{AB}(U\otimes V))|^2 d\mu(U)d\mu(V)\\
&&=\frac1{d_Ad_B} \sum_{i,j} \lambda_i\lambda_j
\mathrm{Tr}((Y_i\otimes Y_j)^\dagger F_{AA} (Y_i\otimes Y_j)F_{BB}),
\end{eqnarray*}
where $F_{AA}$ is the swap operator on
$\mathbb{C}^{d_A}\otimes\mathbb{C}^{d_A}$, $F_{BB}$ is the swap
operator on $\mathbb{C}^{d_B}\otimes\mathbb{C}^{d_B}$. Taking
orthonormal base $|\mu\rangle$ and $|m\rangle$ of $\mathbb{C}^{d_A}$
and $\mathbb{C}^{d_B}$, respectively, gives rise to $F_{AA} =
\sum^{d_A}_{\mu,\nu=1} |\mu\nu\rangle\langle\nu\mu|,~~F_{BB} =
\sum^{d_B}_{m,n=1}|mn\rangle\langle nm|. $ By substituting both
these operators into the above expression, it follows that
\begin{eqnarray*}
&&\int_{\mathrm{U}(d_A)}\int_{\mathrm{U}(d_B)}
|\mathrm{Tr}(\rho_{AB}(U\otimes V))|^2 d\mu(U)d\mu(V)\\
&&= \frac1{d_Ad_B} \sum_{i,j} \lambda_i\lambda_j \mathrm{Tr}(Y_i
Y^\dagger_j)\mathrm{Tr}(Y^\dagger_i Y_j),
\end{eqnarray*}
which is equal to $\frac1{d_Ad_B}\mathrm{Tr}(\rho^2_{AB})$.

Therefore, having access to local
subsystem under local measurement, one can define the measurement disturbance and the change
in the interference visibility can be related to this quantum correlation.


\section{ Visibility and quantum correlation with noisy measurement
disturbance}

In previous section, we have shown that on the
average the change in the visibility of the density operator under
going unitary evolution before and after a complete measurement,  is
directly related to the quantum correlation called the measurement
disturbance. We can ask, how robust is our result to noisy
measurements. Specifically, suppose we define a quantum correlation
measure based on the noisy measurement which causes partial collapse
of the subsystem. Can we still detect the quantum correlation with
change in the visibility? We can model a noisy measurement in which,
with probability $\epsilon$, the complete measurement is performed
and with a probability $(1-\epsilon)$ no measurement occurs. A
complete measurement is the special case of the noisy measurement,
when $\epsilon =1$. This can also be thought of as a quantum channel
$\Phi_{(\epsilon)}$ whose action is defined as
\begin{eqnarray}
\Phi_{(\epsilon)}(\rho_{AB}) := \epsilon \Phi(\rho_{AB}) + (1-
\epsilon) \rho_{AB},
\end{eqnarray}
where $\epsilon \in [0, 1]$ and
$\mathrm{Tr}[\Phi_{(\epsilon)}(\rho)] = 1$. This channel is nothing
but a convex combination of a channel that keeps the state
undisturbed and a channel that implements the strong measurement.
Here, $\Phi_{(\epsilon)}$ may act on one subsystem or both the
subsystems. The noisy measurement can arise in the case of imperfect
detectors and in the case of partial measurements.
We will show that by comparing
the visibility of the density operator undergoing unitary evolution
before and after a measurements that may give incomplete
information, on the average we can detect quantum correlation.

Now for the noisy measurement, we define the quantum correlation measure as
\begin{eqnarray}
Q_{(\epsilon)}(\rho_{AB}) := \|\rho_{AB} - \Phi_{(\epsilon)}(\rho_{AB})\|_\mathrm{HS}.
\end{eqnarray}

From the definition, we can see that the noisy-measurement
disturbance quantum correlation is related to the ideal measurement
disturbance quantum correlation as $Q_{\epsilon}(\rho_{AB})=
\epsilon Q(\rho_{AB})$. If $\epsilon < 1$, then noisy-measurement
disturbance quantum correlation is less than the measurement
disturbance quantum correlation. Below, we will show that for any bipartite
density operator, the difference between the unitary average of the
visibility before and after a noisy measurement is directly related
to $Q(\rho_{AB})$, i.e.,
\begin{eqnarray}
&&\int \left[ |\mathrm{Tr}(\rho_{AB} U)|^2 - |\mathrm{Tr}(\Phi_{(\epsilon)}(\rho_{AB}) U)|^2\right] d\mu(U)\nonumber\\
&&=  \epsilon \left(2-  \epsilon \right) \frac{1}{d_Ad_B} Q(\rho_{AB})^2.
\end{eqnarray}
This shows that our method of witnessing quantum correlation with
visibility is robust against noisy measurements. As long as the
composite system is disturbed by measurement (whether by a complete
von Neumann measurement or noisy measurement), by looking at the
difference of the visibility of the density operator before and
after the measurement, averaged over generic unitary operators, we
can capture the non-classical correlation beyond entanglement.


\section{ Visibility as Witness of Entanglement of pure and mixed states}

One of the prime area of research in quantum information is how to detect entanglement in
composite quantum systems. Once we make sure that the states at disposal are indeed
entangled then we can use them for various quantum information processing tasks.
Here, we will show that for pure bipartite entangled states, having access
to a local system one can detect entanglement in interferometry
visibility. Consider a general pure entangled state
\begin{align}
|\Psi\rangle_{AB} = \sum_i \sqrt{\lambda_i} |\psi_i\rangle_A
\otimes |\phi_i\rangle_B,
\end{align}
where $\lambda_i$ are the Schmidt
coefficients,  and $ |\psi_i\rangle_A \in {\cal H}_A, |\phi_i\rangle
\in {\cal H}_B$ are the local Schmidt bases. Let $|\Psi\rangle_{AB}
\rightarrow (U \otimes I) |\Psi\rangle_{AB}$, and this induces a
local unitary evolution for the subsystem $A$, i.e., $\rho_A
\rightarrow U \rho_A U^{\dagger}$. Now, the interference visibility
for the subsystem $A$ averaged over unitary group is given by
\begin{align}
\int_{\mathrm{U}(d_A)}|\mathrm{Tr}(\rho_A U)|^2d\mu(U) =
1/d_A\mathrm{Tr}(\rho_A^2).
\end{align}
 For pure bipartite entangled state, we
can consider the linear entropy $S_L(\rho_A)= S_L(\rho_B) = E(\Psi)$
as a measure of entanglement, where $E(\Psi) =
1-\mathrm{Tr}(\rho_A^2)$. Hence, the pure state entanglement measure
and the unitary average of the (squared) interference visibility are
directly related. This can be neatly expressed as
\begin{eqnarray}
E(\Psi)= \left(1 - d_A \int |\mathrm{Tr}(\rho_A U)|^2d\mu(U)\right).
\end{eqnarray}
If we use the concurrance $C(\Psi)$ as a measure of entanglement
\cite{woo,rung}, where $C(\Psi) = \sqrt{2(1-
\mathrm{Tr}(\rho_A^2))}$, then the unitary average of the (squared)
interference visibility can be expressed as $C(\Psi)^2= 2\left(1 -
d_A \int |\mathrm{Tr}(\rho_A U)|^2d\mu(U)\right)$.

In the actual experiments, one need not run the interference over all unitaries. Because of our result in section I, 
we have the unitary average as a swap operator that one needs to apply on two copies of
the same system. This is consistent with earlier method of direct detection of entanglement and measurement of
linear and non-linear functions of density operators \cite{ekert}.


For mixed states, via the convex-roof construction, we can define
the entanglement measure (entanglement of formation)
\cite{chb1,chb2}. It is hard to find analytic expressions for the
entanglement of formation and hence lower and upper bounds are very
useful \cite{ter,wer,caves}. For two-qubits  this is a monotonically
increasing function of the concurrence, and it is possible to have a
closed formula using the concurrence \cite{woo}. However, in general
it is not easy. We will give a new lower bound for the entanglement
of formation in terms of the average visibility. Consider a mixed state with
pure state decomposition
\begin{eqnarray}\label{eq:optimal}
\rho_{AB} = \sum_j p_j |\Psi_j\rangle\langle\Psi_j|
\end{eqnarray}
with $\sum_j p_j =1$ and $|\Psi_j\rangle$ are not orthogonal in general.

In quantum information the linear entropy is an useful concept and it has been 
exploited in the monogamy of  concurrence and concurrence of assistance for
multi-particle systems \cite{sband}.
If we use the linear entropy and its convex-roof generalization as a
measure of entanglement of $\rho_{AB}$, then we have the linear entanglement of
formation defined as 
\begin{eqnarray}\label{eq:}
E_F(\rho_{AB}) = \min_{\{p_j, \Psi_j\}} \sum_jp_j E(\Psi_j),
\end{eqnarray}
where the minimum is taken over all decomposition of $\rho_{AB}$,
$E(\Psi_j) = 1- \mathrm{Tr}(\rho_{Aj}^2)$ and $\rho_{Aj} =
\mathrm{Tr}_B(|\Psi_j\rangle\langle\Psi_j|)$.
From the definition, the linear entanglement of formation satisfies
\begin{eqnarray}\label{ef}
 E_F(\rho_{AB}) \leqslant \sum_jp_j (1-\mathrm{Tr}(\rho^2_{Aj})).
 \end{eqnarray}
Now let $\rho_{AB}$ undergoes a local unitary evolution, where the
unitary acts on the subsystem $A$, i.e., $\rho_{AB}\to\rho'_{AB} =
(U\otimes I_B)\rho_{AB}(U^{\dagger}\otimes I_B)$. Then, the visibility
function for $\rho_{AB}$ under the local unitary can be defined as
$V^2(\rho_{AB},U\otimes I_B) = |\mathrm{Tr}(\rho_A U)|^2$, which is
no more than $\sum_jp_j V^2(\rho_{Aj},U)$. By denoting
$\overline{V}^2 = \int_U V^2 d\mu(U)$, the above inequality reads as
$ \overline{V}^2 \leqslant \sum_jp_j \overline{V}^2_j$ with
$\overline{V}^2_j = \int_U  |\mathrm{Tr}(\rho_{Aj} U)|^2 d\mu(U)$.
Therefore, from (\ref{ef}), we obtain
\begin{eqnarray}
\label{ef1}
 E_F(\rho_{AB}) \leqslant 1- d_A\overline{V}^2.
\end{eqnarray}
Thus, the linear entanglement of formation for bipartite mixed state is
upper bounded by a quantity that depends on the average interference
visibility. Looking at (\ref{ef1}), one can also interpret this as a complementary
relation between the entanglement of formation and the average visibility.
In any bipartite state if the average visibility of the local subsystem undergoing unitary evolution
is less then they will share more entanglement. In the subsequent section, we dwell on this
complementarity relation more and reveal a new kind of complementarity between
entanglement and the measurement induced disturbance.


\section{ Complementarity of entanglement and measurement disturbance}

For pure bipartite states, we know that entanglement
measures and quantum correlation measures based on the measurement
disturbance coincide. However, for mixed state
there is no quantitative connection between quantum entanglement and quantum
correlation that supposedly captures something beyond entanglement.
The only relation that we know is the Koashi-Winter relation \cite{kw} that connects
the entanglement of formation and quantum correlation across different
partitions of a tripartite density operator $\rho_{ABC}$.
Here, we will show that for any bipartite mixed state $\rho_{AB}$ the
linear entanglement of formation and quantum correlation respects a
complementarity relation for a given purity of the subsystem state. This
is first ever direct quantitative connection between the entanglement
and the quantum correlation for any bipartite state across the same partition.

Let us consider the quantum correlation measure based on the local
measurement performed on the subsystem $A$. If $\{\Pi_i^A \}$ is a
set of complete one-dimensional orthogonal projectors on ${\cal
H}_A$,  then after the measurement process the density operator
transforms as
\begin{align}
\rho_{AB} \rightarrow \rho_{AB}' &= \Phi(\rho_{AB})=
 \sum_{i} (\Pi_i^A \otimes I^B) \rho_{AB} (\Pi_i^A \otimes I^B) \nonumber\\
& = \sum_i p_i |i\rangle\langle i | \otimes \rho_{Bi}.
\end{align}
Note that for any bipartite density operator $\rho_{AB}$ we have
the relation between the global and the local purity as given by
\cite{ren}
$$
d^{-1}_B \leqslant
\frac{\mathrm{Tr}(\rho^2_{AB})}{\mathrm{Tr}(\rho^2_{A})} \leqslant
d_B
$$

Using (\ref{ef1}) and the above equation
we can derive the complementarity relation. First, note that for the
post-measured state, we have $\mathrm{Tr}(\rho'^2_{AB})  \geqslant
\mathrm{Tr}(\rho'^2_B)/d_A = \mathrm{Tr}(\rho^2_B)/d_A $ (as local
measurement on the subsystem $A$ does not change the purity of $B$). Therefore,
we have the following inequality
\begin{eqnarray}
E_F(\rho_{AB}) +   \frac{1}{d_B} Q(\rho_{AB}) + \frac{1}{d_A d_B}
P(\rho_{B}) \leqslant 1,
\end{eqnarray}
where  $P(\rho_{B}) =  \mathrm{Tr}(\rho_{B}^2) $ is the purity of
the subsystem $B$. This shows that the total quantum
correlation (the entanglement of formation plus the measurement disturbance)
and the subsystem purity can be complementary to each other. If in
the bipartite state, the subsystem state is of less purity, then
the total quantum correlations can be more in that state. For separable
states, we have similar complementary relation between the subsystem
purity and the measurement disturbance. Another consequence of our
new relation is that for a fixed purity of the subsystem state $\rho_{B}$, the
above relation shows yet another kind of complementarity, namely,
there can be tradeoff between pure non-local quantum correlation
such as the linear entanglement of formation and the non-classical correlation such as the
measurement disturbance. This suggests that the entanglement measure and the
non-classical correlation based on measurement disturbance can be genuinely different in nature.
Whether such complementarity relation is a generic feature of entanglement and other non-classical correlations
is an open question. This deserves further exploration in future.



\section{ Conclusions} In quantum theory, interference and
non-classical correlations are two fundamental concepts. In the era of quantum
information they play major roles in information processing tasks.
Whether it is about speed-ups in quantum computer or advantages in
quantum communications one always looks for these non-classical
effects to harness the true power of complex quantum systems. In
this paper, we have shown that these two concepts are
connected in a quantitative manner. We have shown that the
difference in the squared visibility for the density operator before
and after a complete measurement, averaged over all unitary
evolutions, is directly related to the
quantum correlation measure based on the measurement disturbance.
For pure bipartite states, having access to one subsystem, the
unitary average of the visibility gives the concurrence. For mixed
states we have shown that the linear entanglement of formation is bounded
from above by the average of interference visibility. Furthermore, we have
shown that for a fixed purity of the subsystem state, there is a
complementarity relation between the entanglement of formation and the
measurement disturbance. This brings out a fundamental difference
between these two kinds of quantum correlations, hitherto unnoticed.
Our proposal can be tested in interference experiments and the presence
of quantum correlation and entanglement can be revealed.\\~\\

{\it Acknowledgement.---}LZ is grateful for financial support from
Natural Science Foundations of China (No.11301124). AKP thanks
Department of Mathematics, Zhejiang University for supporting his
visit as a K. P. Chair Professor during which this work is carried
out. JW is also supported by Natural Science Foundations of China
(11171301 and 10771191) and the Doctoral Programs Foundation of
Ministry of Education of China (J20130061).



\end{document}